\documentclass[twocolumn,prl,showpacs]{revtex4}

\usepackage{graphicx}
\usepackage{rotating}
\usepackage{amsmath}
\usepackage{amsfonts}
\usepackage{amssymb}
\usepackage{enumerate}
\usepackage{longtable}
\usepackage{dcolumn}
\usepackage{bm}

\begin{document}
\newcommand{\sm}{${\rm SmOFeAs}$}
\newcommand{\smx}{${\rm SmO_{1-x}F_{x}FeAs}$}
\newcommand{\be}{\begin{equation}}
\newcommand{\ee}{\end{equation}}
\newcommand{\bn}{\begin{eqnarray}}
\newcommand{\en}{\end{eqnarray}}

\title{Normal State Correlated Electronic Structure of Iron Pnictides}

\author{L. Craco,$^1$ M. S. Laad,$^2$ S. Leoni,$^1$ and H. Rosner$^1$}
\affiliation{$^1$Max-Planck-Institut f\"ur Chemische Physik fester Stoffe,
01187 Dresden, Germany \\
$^2$Max-Planck-Institut f\"ur Physik komplexer Systeme,
01187 Dresden, Germany }

\date{\rm\today}

\begin{abstract}
We describe the correlated electronic structure of a prototype
Fe-pnictide superconductor, \smx, using LDA+DMFT. Strong, multi-orbital
electronic correlations generate a low-energy pseudogap in the
undistorted phase, giving a bad, incoherent metal in qualitative agreement
with observations. Very good semi-quantitative agreement with the
experimental spectral functions is seen, and interpreted, within a
correlated, multi-orbital picture. Our results show that Fe-pnictides
should be understood as low-carrier density, incoherent metals, in 
resemblance to the underdoped cuprate superconductors.
\end{abstract}


\maketitle

Discovery of high-$T_{c}$ superconductivity (HTSC) in the Fe-based
pnictides~\cite{[1]} is the latest among a host of other, ill-understood
phenomena in $d$-band oxides. HTSC in Fe-pnictides emerges upon doping
a bad metal with spin density wave (SDW) order at ${\bf q}=(\pi,0)$. 
Preliminary experiments indicate~\cite{[2],[3]} unconventional SC. 
Existent normal state data indicate a ``bad metal'' without Landau Fermi 
Liquid (FL) quasiparticles at low energy~\cite{[1]}. These observations 
in Fe-pnictides are reminiscent of cuprate SC. The small carrier density 
(giving rise to carrier pockets), along with Uemura scaling from 
$\mu$-SR~\cite{[4]} similar to hole-doped cuprates strongly suggests 
a SC closer to the Bose condensed, rather than a BCS ($\xi\simeq 1000a$) 
limit.

LDA studies show that SC in Fe-pnictides is associated with the Fe-$d$
states hybridized with As-$p$ states: this leads to two hole, and two
electron-like pockets~\cite{[5]}. Finding of van-Hove singularities and
peaks at ${\bf q}=(\pi,0)$ in the {\it bare} spin susceptibility indicates
a SDW state~\cite{[6],[yild]}, found within Hartree-Fock random phase 
approximation (RPA) studies of {\it effective}, 
two- and four-orbital Hubbard models, in apparent agreement with inelastic
neutron scattering results~\cite{[7]}. The observation of quasi-linear
temperature ($T$) dependence of the resistivity, a pseudogap in
optics~\cite{[8]}, and a spin {\it gap} in NMR~\cite{[nmr]}, however, 
is a benchmark feature of the relevance of strong, dynamical spin and 
charge correlations in the pnictides. In cuprates, these are seen in the 
{\it underdoped} state in the proximity of a Mott insulator (MI), suggesting
that the Fe-pnictides might be closer to a MI than generally
thought~\cite{[9]}. Actually, the undoped pnictides show an insulator-like
resistivity {\it without} magnetic order for $T>150$~K~\cite{[1]}. Onset
of bad metallic behavior correlates with a structural
(tetragonal-orthorhombic (T-O)) distortion at $T^{*}\simeq 150$~K,
{\it below} which SDW order sets in. The small carrier number apparently
generated upon the structural distortion accords with the observed high
resistivity, lending further credence to such a view.

The above suggests that one should study a single Fe-As layer with
strong electronic correlations to begin with.  Here, we study the
five-orbital Hubbard model within the local-density-approximation plus
dynamical-mean-field-theory (LDA+DMFT) approach, incorporating  
one-electron band structure aspects. Extant LDA+DMFT works give either
a strongly renormalized FL~\cite{[10]} or an orbital selective, incoherent,
pseudogapped metal~\cite{[11]}. Apart from the known sensitivity to the
value of $J_{H}$ (Hund coupling), the LDA+DMFT spectra show noticeable
qualitative differences at low energy. Are Fe-pnictides then strongly
renormalized FL metals, or incoherent non-FLs in their normal state?
Comparison with experimental one-particle spectra should go a long way
toward resolving this question. We do this in this work.

Photoemission (PES) and X-ray absorption (XAS) studies are a reliable tool
to study the {\it correlated} electronic structure of $d$- and $f$-band
compounds~\cite{[12]}. To date, the only PES experiments have been performed
on \smx~\cite{[13]} and ${\rm LaO_{1-x}F_xFeAs}$~\cite{ishida}. Very 
recently, XAS has also been performed for \smx~\cite{[14]}. Together, they 
provide additional evidence for the ``incoherent metal'' normal state in 
Fe-pnictides. The PES spectra show a kink at low energy, 
$\Omega=15$~meV~\cite{[13]}, below the T-O distortion followed by SDW 
order. This kink sharpens with cooling, and evolves, apparently smoothly, 
across $T_{c}$. Its microscopic origin is an enigma. Is it related to the 
T-O distortion, or to the SDW transition? Is it observed only for 
electron-doped systems? Correspondingly, XAS shows a well-defined peak at 
$0.5$~eV, and a transfer of weight from high energy ($2.0$~eV) to low 
energy with F doping, a characteristic feature of correlated systems. 
Answering these questions within a {\it correlated} electronic structure 
approach provides deeper insight into the underlying correlations in the 
non-FL metal phase, aiding in the search to identify mechanism(s) of 
superconductivity itself.

\begin{figure}[thb]
\begin{center}
\includegraphics[width=\columnwidth]{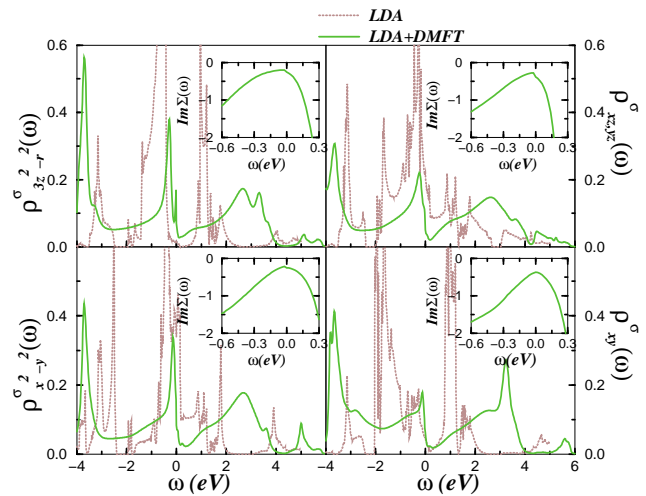}
\end{center}
\caption{(Color online)
Orbital-resolved DOS: LDA and LDA+DMFT ($U=4.0$~eV, $U'=2.6$~eV and 
total band filling, $n$=6.0) for \sm. The insets show the imaginary part 
of the corresponding self-energies, showing clear evidence of non-FL 
behavior.}
\label{fig2}
\end{figure}

Starting with the high-$T$ tetragonal structure with
lattice parameters found in Ref.~\cite{struc}, one-electron band structure
calculations were performed for \sm~using the linear muffin-tin orbitals
(LMTO)~\cite{ok} scheme in the atomic sphere approximation~\cite{sphere}.
LDA provides valuable description of the relevant orbitals on a
single-particle microscopic level, but falls short of describing dynamical
correlations in $d$- and $f$-band compounds. This requires ``marrying'' LDA
to DMFT, which enables direct access to the {\it correlated} spectral
functions~\cite{[15]}. The one-electron part for \sm~is
$H_{0}=\sum_{{\bf k},a,\sigma}
\epsilon_{a}({\bf k})c_{{\bf k},a,\sigma}^{\dag}c_{{\bf k},a,\sigma}\;,$
where $a=x^{2}-y^{2},3z^{2}-r^2,xz,yz,xy$ label the diagonalized, five 
$d$ bands. The corresponding density-of-states (DOS) (Fig.~\ref{fig2}) 
shows that {\it all} the five $d$-bands cross the Fermi energy, $E_{F}$, 
but the $3z^{2}-r^2$ band is almost gapped at $E_{F}$. While the $xy$-band 
has a deep pseudogap at $E_{F}$, the $x^{2}-y^{2},xz,yz$-bands have large
DOS at $E_{F}$. In Fe-pnictides, the $d^{6}$ configuration of Fe$^{2+}$
dictates that the full, multi-orbital (MO) Coulomb interactions must be
included. These constitute the interaction term, which reads

\bn
\nonumber
H_{int}=U\sum_{i,a}n_{ia\uparrow}n_{ia\downarrow}
+ U'\sum_{i,a \ne b}n_{ia}n_{ib}
-J_{H}\sum_{i,a,b}{\bf S}_{ia}.{\bf S}_{ib}\;.
\en
We choose parameters employed by Haule {\it et al.}~\cite{[10]}, $U=4.0$~eV, 
$U'=U-2J_H=2.6$~eV, and $J_{H}=0.7$~eV, along with the five LDA bands, and 
solve $H=H_{0}+H_{int}$ within LDA+DMFT. To solve the MO-DMFT equations, we 
use the MO iterated-perturbation-theory (IPT) as an impurity solver. Though 
not quantitatively exact, it has many advantages. It is numerically very 
efficient, is valid at $T=0$, in contrast to QMC, and self-energies 
$(\Sigma_a(\omega))$ can be extracted very easily. These are of particular 
importance in a complicated MO situation that occurs in Fe-pnictides.

Given orbital-induced anisotropies in the LDA, strong MO correlations
renormalize various $d$-bands in widely differing ways. Generically, one
expects these to partially (Mott) localize a subset of $d$-bands, leading
to orbitally selective Mott transitions (OSMT), and bad metallic
states~\cite{[16]}. This requires strong $U,U'$. Within LDA+DMFT, this
orbital selective mechanism involves two renormalizations: $(a)$ static
(MO Hartree) shifts rigidly move various $d$-bands relative to each other
by amounts depending upon their on-site orbital energies and occupations,
and, more importantly, $(b)$ dynamical effects of strong $U,U',J_{H}$
drive large spectral weight transfer (SWT) over wide energy scales. Upon
small changes in {\it bare} LDA parameters, large {\it changes} in SWT
lead to OSMT, as well as to incoherent metallic phases characteristic
of a wide variety of correlated systems. With parameters for Fe-pnictides
as above, does an OSMT occur~\cite{[11]}, or does it not~\cite{[10]}?
What is the origin of the observed incoherent metallic behavior?   

\begin{figure}[thb]
\begin{center}
\includegraphics[width=\columnwidth]{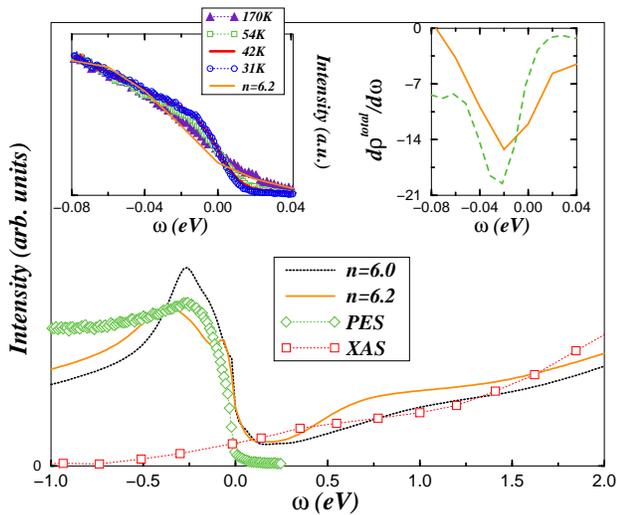}
\end{center}
\caption{(Color online)
Comparison between theoretical (LDA+DMFT DOS) and experimental (PES, XAS)
spectra in the nFL metallic phase of \smx, ${\rm x}=0.15$. Notice the good
agreement with experimental PES result up to 0.5~eV binding energy. PES
and XAS data are taken, respectively, from Refs.~\cite{[13]} and~\cite{[14]}.
The insets show the $T$-dependence of the experimental angle-integrated PES 
(left) at low binding energies for $x=0.15$, and the derivative of the total 
DOS (for $n=6.2$).  The latter show a sharp peak at $\Omega=20.0$~meV 
(solid: theory, dashed: experiment), indicating the kink structure in very 
good agreement with the experimental value of 15~meV below $T^{\star}$.}
\label{fig3}
\end{figure}

To answer these questions, we now turn to our results. In sharp contrast
to LDA, our LDA+DMFT results show drastic modification of the spectral
functions. The dynamical correlations lead to dramatic spectral weight
redistribution over large energy scales $O(5.0)$~eV. Most interestingly,
we find no FL quasiparticle signatures in the low-energy spectra; instead,
the metallic state is totally {\it incoherent}. In contrast to earlier
work~\cite{[11]}, this occurs even {\it without} a strict orbital selective
Mott localization, though almost all bands are very close to Mott
localization. The orbital-resolved self-energies (inset in Fig.~\ref{fig2})
clearly reveal this aspect: large damping at $E_{F}$ destroys the FL
quasiparticle, giving an incoherent, pseudogapped, bad metallic state.
This incoherent state has also been found in previous LDA+DMFT
works~\cite{[10],[11]}, and, as discussed there, is in qualitative
agreement with experimental observations. Here, however, we also show
that our results are in very good semi-quantitative agreement with key
features of the experimental PES {\it and} XAS spectra, see
Fig.~\ref{fig3}.

Analyzing the LDA+DMFT spectra of Fig.~\ref{fig2}, we find that the
$3z^{2}-r^2,x^{2}-y^{2}$ as well as the $xz,yz$ orbitals continue to
be almost degenerate (the first pair is split by $0.06$~eV). This is also
reflected by the fact that the DMFT spectra of $3z^{2}-r^2,x^{2}-y^{2}$
bands show noticeable similarities, even as, interestingly, large differences
between them exist at level of LDA. Of interest are the sharp, very
low-energy ($20$~meV {\it below} $E_{F}$) structures in the
$3z^{2}-r^2,x^{2}-y^{2}$ spectra. In the light of their near-degeneracy,
we ascribe these peaks to the low-energy orbital fluctuations (coupled
to charge fluctuations) in this two-fold degenerate sector. This is an
interesting manifestation of the two-fold $3z^{2}-r^{2},x^{2}-y^{2}$ orbital
degeneracy surviving in Fe-pnictides, and explicitly requires strong
MO correlations.  Using LDA+DMFT, we have also estimated the spin state 
on Fe sites. With parameters as above, the quantity
$\sqrt{<S_{z,total}^{2}>} \simeq 1.8$, slightly less than the atomic
value of $S=2$, and consistent with a high-spin state. The observed
suppression of the sublattice magnetization in neutron scattering should 
therefore be attributed to strong geometric frustration in Fe pnictides, 
as discussed recently by several authors~\cite{[9],[yild]}. This is hitherto
an open issue: should Fe-pnictides be modelled using $S=1$ or $S=2$ 
models~\cite{[9]}? Our analysis resolves this issue in favor of $S=2$ 
modelling.  We note that this implies strong $J_{H}$, and, from $d$-shell 
quantum chemistry, larger $U,U'$, putting the pnictides into the
``strongly correlated'' class of materials.

\begin{figure}[thb]
\begin{center}
\includegraphics[width=\columnwidth]{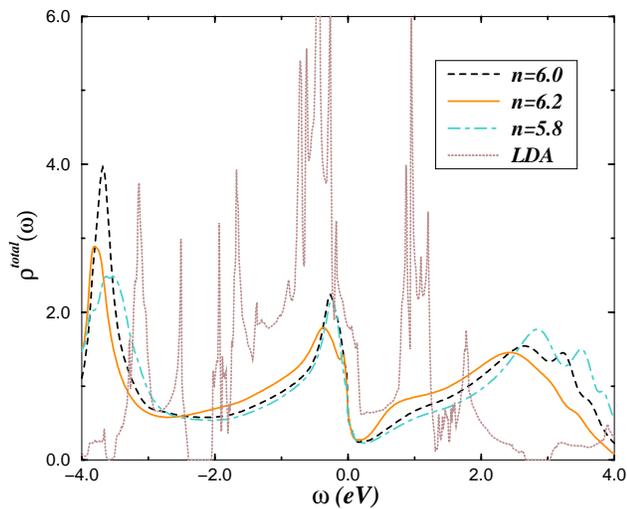}
\end{center}
\caption{(Color online)
Total LDA+DMFT spectral functions for \smx. Notice the large dynamical 
SWT upon electron and/or hole doping. LDA result for \sm~is shown for 
comparison.}
\label{fig4}
\end{figure}

An LDA+DMFT-experiment comparison, carried out here for the first time for
Fe-pnictides, puts these features into deeper perspective. In
Fig.~\ref{fig3}, we compare our results to recent PES and XAS results on
\smx~with ${\rm x}=0.15$. Very good semi-quantitative agreement with
experiment is visible. In particular, the broad peak at $-0.3$~eV
in PES, as well as that at $0.5$~eV in XAS, are in nice agreement with our
calculations. Further, the transfer of spectral weight from the $2.0$~eV
feature to that at $0.5$~eV in XAS is also reproduced semiquantitatively
in the correlated spectra. Repeating the DMFT calculations with $U=5.0$~eV,
as in Ref.~\cite{[10]}, we found worse agreement with {\it both} PES
and XAS data. Most importantly, PES reveals a kink at $15$~meV upon cooling
the sample below $T^{*}=150$~K, where a T-O distortion, {\it followed}
by the ${\bf q}=(\pi,0)$ SDW order, takes place. This feature sharpens with
decreasing $T$, and weakens with doping, but does not undergo further
change across the SC $T_{c}$.  This appears to be a more generic feature of
the electronic structure of Fe-pnictides, as similar evolution of the low
energy pseudogap has also been resolved in
${\rm LaO_{1-x}F_{x}FeAs}$~\cite{ishida}. Analysing our LDA+DMFT spectra 
for $n=6.2$ ($n$ is the total band filling of the $d$ shell), we find a 
sharp nonanalytic structure in d$\rho^{total}(\omega)$/d$\omega$ at an 
energy of $20$~meV, implying a kink in the DOS at that energy, in very 
good semi-quantitative agreement with PES results. Further, analyzing the 
orbital-resolved DOS, this feature is seen to originate from the 
$3z^{2}-r^2,x^{2}-y^{2}$ (DMFT) spectra. Since this structure is a 
consequence of near two-fold ($3z^{2}-r^2,x^{2}-y^{2}$) orbital degeneracy, 
as discussed above, its appearance below $T^{*}$ now has an attractive 
interpretation: it reflects the low energy, coupled charge-orbital 
fluctuations in this pnictide. It appears only below $T^{*}$ because the 
T-O distortion, interpreted as a Jahn-Teller instability, occurs at 
$T^{*}$~\cite{[17]}, and lifts this degeneracy. The itinerant (albeit 
incoherent) character of the system suppresses the bare 
$3z^{2}-r^2,x^{2}-y^{2}$ splitting to small values ($20$~meV, as pointed 
out above).  Strong orbital fluctuations in this almost degenerate orbital 
sector coupled to one-electron Green functions enter the {\it dynamical}, 
second order contributions (of the generic form $\int G_{a}^{(0)}(\omega-\omega_{1}-\omega_{2})G_{b}^{(0)}(\omega_{2})G_{a}^{(0)}(\omega-\omega_{1})d\omega_{1} d\omega_{2}=\int \chi_{ab}^{(0)}(\omega_{1})G_{a}^{(0)}(\omega-\omega_{1})d\omega_{1}$,
with $\chi_{ab}^{(0)}(\omega)$ the inter-orbital susceptibility describing
{\it dynamical} orbital correlations) to the self-energies, and hence show
up in the LDA+DMFT spectra as sharp, low energy peaks in the
orbital-resolved spectral functions, as seen in our results. We have also 
computed the LDA+DMFT spectra for hole doping ($n=5.8$, dot-dashed line in
Fig.~\ref{fig4}). In contrast to electron doping, no noticeable change is
observed in the {\it low energy} spectra; we predict that PES/XAS on 
hole-doped Fe-pnictides will show this in future.   

Since the kink in PES is now related to the T-O distortion, rather than the
SDW, it should be smeared out for $T>T^{*}$, exactly as observed. Therefore, 
its survival without apparent modification across $T_{c}$ is not connected to
the destruction of SDW order apparently required for SC to emerge. {\it If}
this turns out to be generic for Fe-pnictides, it would imply an
{\it indirect} link, at most, to SC, to the extent that it reflects
electronic structure changes (viz, removal of $3z^{2}-r^2,x^{2}-y^{2}$
orbital degeneracy) {\it required} for the SC instability to emerge from
such a normal state.  ``Melting'' of the T-O distortion upon F doping
implies rapid suppression of this kink feature in our picture: this is
indeed seen in our LDA+DMFT spectra for $n=6.2$ (orange line) curves. Only 
a weak remnant of the $15$~meV kink is resolved at $n=6.2$, as shown in 
the right inset of Fig.~\ref{fig3}, indicative of surviving {\it uncorrelated} 
distortions among an ``undoped'' fraction of Fe-sites, even as long-range 
distortion melts with doping.

Based on this detailed theory-experiment agreement, we discuss the
implications of our work on SC.  First, since {\it all} $d$ bands cross
$E_{F}$, SC should involve inducing the SC gap on all FS sheets. This 
does not necessarily conflict with two-band Hubbard model 
results~\cite{[raghu2]}, since the multi-band SC proximity 
effect~\cite{[rice]} could operate here. Once SC pairing occurs in the 
$d_{xz,yz}$ manifold, such an effect could induce secondary gaps over 
the remaining FS sheets. Second, our results indicate that having 
$3z^{2}-r^2,x^{2}-y^{2}$ orbital degeneracy drives a Jahn-Teller T-O 
distortion, to the detriment of SC, as seen.  Finally, our finding of 
a strongly incoherent ``normal'' state with drastically reduced charge carrier 
number (given the proximity to a Mott insulator in DMFT) is consistent 
with observations~\cite{[1],[2],[4]} indicating a non-FL metal 
with low carrier density SC. Many of these observations are reminiscent 
of those seen in HTSC cuprates~\cite{[24]} up to optimal doping, putting 
the Fe-pnictides into the ``strongly correlated, unconventional, HTSC'' 
category.

In conclusion, using a correlated electronic structure (LDA+DMFT) approach, 
we show that strong dynamical correlations are essential to proper
understanding of the basic physics of Fe-pnictides.  As in cuprates,
the normal state that becomes unstable to unconventional~\cite{[2],[3],[4]}
SC is {\it not} a Fermi liquid. Our LDA+DMFT result gives an incoherent,
non-FL state with small carrier density, as observed 
experimentally~\cite{[4]}. Very good semi-quantitative agreement with 
extant PES/XAS data lend strong credence to our view of Fe-pnictides as 
multi-orbital, strongly correlated materials close to the 
itinerant-localized boundary.

L.C. thanks A. Ormeci for discussions.
L.C. and H.R. thank the Emmy-Noether Program of the DFG for support.
M.S.L. thanks the MPIPKS for financial support.

\end{document}